\documentclass[fleqn,usenatbib]{mnras}
\usepackage{mathptmx}
\usepackage[T1]{fontenc}
\DeclareRobustCommand{\VAN}[3]{#2}
\let\VANthebibliography\thebibliography
\def\thebibliography{\DeclareRobustCommand{\VAN}[3]{##3}\VANthebibliography}
\usepackage{graphicx}	
\usepackage{amsmath}	
\usepackage{amssymb}	
\usepackage{rotating}
\usepackage{sidecap}
\usepackage{lscape}
\usepackage{multicol, blindtext}

\usepackage{aecompl}
\usepackage{wasysym}
\usepackage{array}
\usepackage{times}
\usepackage{url}
\usepackage{xcolor}
\usepackage{hyperref}
\hypersetup{colorlinks=true,citecolor=blue,linkcolor=blue,filecolor=blue,urlcolor=blue}

\newcommand{\msun}{{\,\rm M_\odot}}

\newcommand{\kms}{\,{\rm km}\,{\rm s}^{-1}}

\newcommand{\kpc}{\,{\rm kpc}}

\newcommand{\Mpc}{\,{\rm Mpc}}

\def\prl{Phys. Rev. Lett.}

\def\jcap{J. Cosmol.  Astropart. Phys.}
\def\aap{A\&A}
\def\apj{ApJ}

\def\apjl{ApJ}
\def\mnras{MNRAS}

\def\aj{AJ}

\def\na{New Astronomy}

\def\physrep{Phys. Rep.}
\def\nat{Nature}

\def\apjs{ApJS}
\def\prd{Phys. Rev. D}

\newcommand{\comment}[1]{}

\title[Endothermic SIDM primordial particle states]
    {Varying primordial state fractions in exo- and endothermic SIDM simulations of Milky Way-mass haloes}
\author[A.~Leonard et al.]
    {{Aidan Leonard$^{1}$ \thanks{E-mail: aidanl@mit.edu},
    Stephanie O'Neil$^{1}$,
    Xuejian Shen$^{1}$,
    Mark Vogelsberger$^{1,2}$,}
    \and
    {Olivia Rosenstein$^{1}$,
    Haotian Shangguan$^{3}$,
    Yuanhong Teng$^{4}$ and
    Jiayi Hu$^{5}$}\\
    $^{1}$Department of Physics and Kavli Institute for Astrophysics and Space Research,
           Massachusetts Institute of Technology,
           Cambridge, MA 02139, USA\\
    $^{2}$The NSF AI Institute for Artificial Intelligence and Fundamental Interactions, Massachusetts Institute of Technology, Cambridge, MA 02139, USA\\
    $^{3}$Department of Computer Science, Boston University, Boston, MA, 02215\\
    $^{4}$University of Science and Technology of China, Hefei, Anhui 230026, China\\
    $^{5}$Department of Mathematics, Columbia University, New York, NY 10027, US
    }

\begin{document}

\date{Accepted 2024 May 11. Received 2024 May 03; in original form 2024 January 24}

\pagerange{\pageref{firstpage}--\pageref{lastpage}}
\pubyear{2023}

\maketitle

\label{firstpage}

\begin{abstract}
Self-interacting dark matter (SIDM) is increasingly studied as a potential solution to small-scale discrepancies between simulations of cold dark matter (CDM) and observations.
We examine a physically motivated two-state SIDM model with both elastic and inelastic scatterings.
In particular, endothermic, exothermic, and elastic scattering have equal transfer cross sections at high relative velocities ($v_{\rm rel}\gtrsim400~{\rm km/s})$.
In a suite of cosmological zoom-in simulation of Milky Way-size haloes, we vary the primordial state fractions to understand the impact of inelastic dark matter self-interactions on halo structure and evolution. In particular, we test how the initial conditions impact the present-day properties of dark matter haloes.
Depending on the primordial state fraction, scattering reactions will be dominated by either exothermic or endothermic effects for high and low initial excited state fractions respectively.
We find that increasing the initial excited fraction reduces the mass of the main halo, as well as the number of subhaloes on all mass scales. The main haloes are cored, with lower inner densities and higher outer densities compared with CDM.
Additionally, we find that the shape of the main halo becomes more spherical the higher the initial excited state fraction is.
Finally, we show that the number of satellites steadily decreases with initial excited state fraction across all satellite masses.
\end{abstract}


\section{Introduction}
\label{sec:intro}

Dark matter (DM) forms about 85\% of the mass budget of the Universe, but it has never been observed directly due to its minimal interaction with baryonic matter.
Models for DM are often explored through cosmological simulations, which are then compared to observational constraints - for example, from rotation curves, strong or weak gravitational lensing, and the luminosity function of galaxies.
The prevailing model is the cold DM plus cosmological constant ($\Lambda$CDM) model, where DM is assumed collisionless and has non-relativistic velocities since the early universe.
Rather than committing to a description of DM on the particle physics scale, CDM is flexible and uses macroscopic properties that apply to a broad class of DM candidates.
Weakly interacting massive particles (WIMPs) have been a particularly promising candidate for CDM since the freeze-out density naturally falls within the energy range of the weak force \citep[e.g.][]{Lee1977}, but this class of particles is becoming increasingly constrained.

CDM is successful in describing the large ($\gtrsim 1$ Mpc) scale structure of the universe.
Simulations employing CDM have recreated, for example, the observed distribution of galaxies and halo mass function implied through abundance matching \citep[e.g.][]{Springel2008,BoylanKolchin2009,Klypin2011,Garrison-Kimmel2014a}.
However, there are several small-scale discrepancies between observational data and the predictions of CDM \citep{Bechtol2022}.

Historically, the main discrepancies include the core-cusp, diversity of rotation curves, missing satellites, and too-big-to-fail problems.
The core-cusp problem refers to the observation that satellites have a cored inner density profile $\rho\propto r^n$  with $n\approx0$ but simulations predict a cuspy inner profile with $n\approx-1$ \citep{Marchesini2002,deBlok2008,KuziodeNaray2008,Walker2011}.
Additionally, observed galaxies vary in their inner densities  in both field \citep{Oman2015,Creasey2017} and satellite \citep{Zavala2019b,Hayashi2020} populations while simulated satellites are much more uniformly described by the NFW profile.

The missing satellites problem refers to the historical discrepancy where many more satellites were predicted in simulations than satellite galaxies in observations \citep{Moore1999,Springel2008,Klypin2011,Garrison-Kimmel2014a}.
More recently, however, more faint dwarf galaxies have been observed that, combined with accounting for the completeness of surveys, bring the number counts for observed galaxies and simulated haloes into agreement \citep{Kim2018,Drlica-Wagner2020}.
Finally, the too-big-to-fail problem refers to the scenario where there are more large DM haloes predicted than large galaxies seen, but the predicted haloes have a large enough gravitational potential that they should readily form galaxies \citep{Boylan-Kolchin2011,Boyer2012,Garrison-Kimmel2014b}.
The potential should also be strong enough to resist losing the majority of the stars to stripping.

These discrepancies were primarily observed in DM-only simulations, where baryons are ignored.
Because of the high dark-to-baryonic matter ratio, galaxies are expected to form alongside DM overdensities.
However, despite making up only a fraction of the total mass, baryonic matter has a significant influence on galaxy formation.
Feedback from supernovae or stellar winds, for example pushes, DM outwards \citep{Navarro1996b,Governato2012,DiCintio2014a,Dicintio2014b}, which has produced cores in simulations across a range of galaxy sizes \citep{Pontzen2012,Keller2016,Burger2022}.

Many prescriptions for baryonic processes rely on subgrid models, where the relevant physics occurs at scales less than the resolution of the simulation.
The effects of these subgrid models are sensitive to model parameters and can vary significantly and impact key comparisons between simulations and observations \citep{Vogelsberger2020}.
The hydrodynamics solver for gas, e.g. using the moving-mesh scheme \textsc{Arepo} or the smoothed particle hydrodynamics scheme \textsc{Gadget}, can change the size of disks that are produced \citep{Torrey2012}.
The star formation models are also not well understood.
Altering the density threshold for star formation in gas influences whether cores are formed in dwarf galaxies \citet{Benitez-Llambay2019}, altering the primary factor that determines the star formation rate, e.g. a density threshold or an H$_2$-based model, changes the number of satellites by a factor of two \citep{Munshi2019}, and altering the star formation efficiency affects the morphology of galaxies \citep{Agertz2015}.
These parameters can thus be the difference between matching observations or producing unrealistic galaxies.

Although baryonic models may mitigate some of the small-scale discrepancies, it is difficult to resolve them all simultaneously.
Since DM is also not well understood, there is significant interest in modifications to CDM that alleviate its small-scale problems while maintaining its large-scale accuracy \citep{Cyr-Racine2016,Vogelsberger2016}.
In addition, it is also difficult to predict how effects from baryonic physics and alternative DM combine.
Alternative DM models can alter star formation and gas distribution \citep{Shen2024} and the stellar density profile \citep{Vogelsberger2014c}.

Early proposed DM candidates based on known physics are heavily constrained.
Warm DM (WDM), for instance, that may be made up of particles like  neutrinos \citep{Dodelson1993,Shi1998}, are strongly constrained by the Lyman-$\alpha$ forest \citep{Viel2013, Schneider2014a,Irsic2017a,Hooper2022}.
Weakly interacting massive particles (WIMPs), while promising due to their interaction scale, have become increasingly limited in their allowed cross sections as both direct and indirect search efforts have yet to detect them \citep{MarrodanUndagoitia2016,Strigari2013}.
Like with baryonic models, it is difficult to simultaneously solve all the small-scale problems \citep{Colin2008,Polisensky2015,Lovell2017}.

This has led to the increasing interest in self-interacting DM (SIDM) \citep{Spergel2000}, which produces distinct direct detection signals \citep{Vogelsberger2013b}.
This is a large class of particles that allow for self-interactions between DM particles but not between DM and baryons beyond gravity.
With small interaction cross sections, SIDM does not alter the large-scale structure of the universe \citep{Rocha2013,Vogelsberger2016,Sameie2019,Andrade2022} but can alleviate small-scale problems by transferring orbital energy between particles \citep[][and references therein]{Tulin2017}, undergoing core-collapse \citep{Kaplinghat2019}, causing variations in galactic rotation curves \citep{Creasey2017}, and combining with tidal forces \citep{Dooley2016,Sameie2020}.

The parameter space of SIDM still remains largely unconstrained.
Viable constant transfer cross sections per mass ($\sigma_{\rm T}/m$) do not significantly alter the small-scale structure of galaxies \citep{Zavala2013,Zeng2022,Silverman2023}, so recent models have introduced velocity dependence such that the cross section depends on the relative velocity of the scattering particles \citep{Vogelsberger2012b,Meshveliani2022}.
SIDM particles also occur in many DM models, including atomic DM \citep{Cline2013b,Cyr-Racine2012,Boddy2016}, composite DM \citep{Frandsen2011,Boddy2014,Chu2018b}, strongly interacting massive particles \citep{Hochberg2014a,Hochberg2014b,Bernal2015,Choi2017,Hochberg2018b}, charged DM \citep{McDermott2010,Liu2019,Dvorkin2019}, mirror DM \citep{Mohapatra2001,Foot2008,An2010,Chacko2021}, and secluded DM \citep{Boehm2003b,Pospelov2007,Feng2008}.

Additionally, SIDM scattering events can be elastic or inelastic.
While many models initially explored elastic SIDM \citep{Vogelsberger2013b, Rocha2013,Zavala2013,Elbert2015,Andrade2022}, models including inelastic collisions are rising in interest due to their ability to alter the energy within the DM particles \citep{mvm2010,mvm2010b,mvm2014PRL,mvm2014JCAP,Schutz2014,Blennow2016,Zhang2016,Vogelsberger2019,Shen2021,Shen2024a,O'Neil2023} and have important cosmological consequences~\citep[e.g.][]{Fan2013,Foot2014,Vogelsberger2019,Xiao2021,Roy2023}.
One way to provide the mechanism for energy transfer during inelastic reactions is for the DM particles to exist in multiple states \citep[e.g.][]{Schutz2014,Vogelsberger2019,O'Neil2023}.
Other mechanisms, like decay \citep[e.g.][]{Wang2014}, could have similar effects, but the multi-state model provides a large amount of flexibility and is what we focus on in this paper.
The energy difference between the states can then be exchanged with the kinetic energy of the particles.

The parameter space in the multistate model is then defined by the ground state mass ($m_{\chi^1}$), the energy difference between the two states ($\delta$), the dark force mediator mass ($m_\phi$), and the coupling between the mediator and the DM particle ($\alpha$) \citep{Schutz2014}. By adjusting these parameters, a wide range of SIDM behaviours can be achieved, including elastic, exothermic, and endothermic reactions \citep{Vogelsberger2019,O'Neil2023}.
It is not currently feasible to fully explore this parameter space with simulations, so we take representative models that emphasise different reactions.

In addition to the parameters of the SIDM model, it is also necessary to determine the initial abundance of states in the early universe, which depends on the cosmological history of the DM model.
Previous work \citep[e.g.][]{Vogelsberger2019,O'Neil2023} has been restricted to initialising the DM particles entirely in one state or split evenly between the two states.
The initial abundance of each state has marked effects on the resulting SIDM behaviour.
If the particles are initially all in the ground state, for example, a model that is primarily exothermic will be indistinguishable from CDM.
Generically, it is expected that the ground state is energetically favoured by the high-density environment of the early universe \citep{mvm2001a,mvm2010,mvm2014JCAP}.
Depending on the processes that determine the initial abundances, for example freeze-in or resonant annihilation, it is possible to obtain varying amounts of DM in the ground and excited states \citep{Brahma2024}.

In this paper, we expand on the work done in \citet{O'Neil2023}.
Using the described multi-state SIDM framework, they introduced a model where elastic, exothermic, and endothermic scattering events had similar cross sections and therefore all contributed to the SIDM effects in the halo.
They then ran this model, along with several comparison models, in DM-only, Milky Way-sized zoom-in simulations.
In isolation, elastic scattering will produce cores in a halo, exothermic scattering will push particles to larger radii and make it more difficult for satellites to form, and endothermic scattering will lead to high central densities in both Milky Way-mass and satellite haloes.
Thus, in isolation, endothermic scattering is expected to exacerbate small-scale DM discrepancies, but the non-linear interplay between the different reaction types makes it difficult to predict how DM haloes will behave when all reactions are able to occur.
They compared this model to the model in \citet{Vogelsberger2019}, which suppressed the endothermic reactions, and showed that viable DM haloes could be produced even with a significant endothermic cross section.

Here, we explore the effects of altering the initial abundance of DM particle states on the introduced endothermic SIDM model.
\citet{O'Neil2023} explored only the case in which all particles started in the ground state.
We study the effects of initial fractional abundance of each state between 0 and 1 in a suite of eleven Milky Way-mass zoom-in simulations using the same initial conditions as in \citet{O'Neil2023}.
Although not each of these initial abundances is equally likely, it is necessary to explore the intermediate abundances to understand how the halo properties transition.

The paper is organised as follows.
In Section \ref{sec:methods}, we describe our simulations (\ref{sec:methods_sims}) and the SIDM model (\ref{sec:methods_model}).
In Section \ref{sec:results}, we show the results of varying the initial abundances on properties of DM haloes: density profiles (\ref{sec:results_density}), abundances of each state at $z = 0$ (\ref{sec:results_excited_fraction}), halo shape (\ref{sec:results_shape}), velocity dispersion and anisotropy (\ref{sec:results_kinematic}), and substructure (\ref{sec:results_substructure}). We also show the redshift evolution of the haloes (\ref{sec:results_redshift}).
Finally, we summarise our conclusions in Section \ref{sec:conclusions}.
\section{Methods}

\label{sec:methods}

\begin{figure}
    \centering
    \includegraphics[width=\columnwidth]{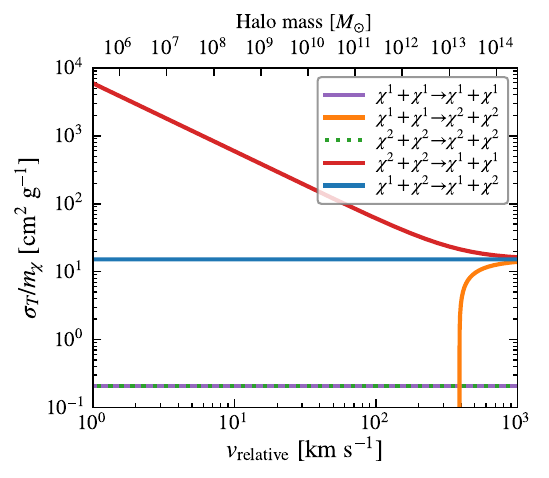}
    \caption{The transfer cross sections as a function of relative particle velocity for each scattering reaction in the endothermic SIDM model. The top axis shows the halo $M_{200,\rm{mean}}$ with circular velocity $v$ at $R_{200,\rm{mean}}$.  Elastic scattering between same state particles is very unlikely, while the elastic scattering between different state particles is similar in likelihood to down- and up-scattering.  Down-scattering is most likely at lower relative velocities, and there is a velocity threshold for up-scattering to ensure that particles have enough kinetic energy to transform to the energy difference between the ground and excited states.}
    \label{fig:cross_sections}
\end{figure}

\subsection{Simulations}
\label{sec:methods_sims}

The analysis in this paper is based on a suite of 11 dark matter-only (DMO) cosmological zoom-in simulations of Milky Way-mass haloes. The simulations are performed using the multi-physics, moving-mesh, hydrodynamic code {\sc Arepo}~\citep{Springel2010,Weinberger2020}, which employs the tree-particle-mesh (Tree-PM) algorithm for gravity with periodic boundary conditions. The main target haloes in the zoom-in region have $M_{\rm200,mean}\approx2\times10^{12}\msun$ at $z=0$, where $M_{\rm200,mean}$ is the mass within the radius where the average density is $200$ times the mean density of the universe.
The simulations use the following cosmological parameters:
matter density fraction $\Omega_{\rm m}=0.302$, dark energy fraction $\Omega_\lambda=0.698$, and Hubble constant $H_0=100h\:{\rm km s}^{-1}{\rm Mpc}^{-1}$ where $h=0.691$.

The main halo and its surroundings are made up of 55,717,200 high-resolution DM particles. 
The high-resolution region is surrounded by particles of mass $1.67\times10^6\:\msun$ and $1.13\times10^8\:\msun$ farther away from the main halo.
For the high-resolution region, the effective Plummer equivalent gravitational softening length of DM is $\epsilon=0.153 \kpc$ and the DM particle mass resolution is $m^{\rm zoom}_{\rm dm} = 2.2 \times 10^{5} \msun$. The background particles get less resolved the farther they are from the main halo in particle masses in the range $m^{\rm back}_{\rm dm} = 1.76\times10^6 - 9.03\times 10^8 \msun$.
The gravitational softening lengths for these particles are adjusted by a factor of $(m^{\rm back}_{\rm dm}/m^{\rm zoom}_{\rm dm})^{1/3}$. The SIDM model is not implemented for the background particles, which behave as CDM.

The periodic simulation box has a side length of $100 \Mpc/h$. The initial conditions are generated at redshift $z=127$ using \textsc{Music} \citep{Hahn2011}. All the simulations in our suite were run using the same initial conditions.

Haloes are identified using a Friends-of-Friends (FoF) algorithm \citep{Davis1985}.
This spatially identifies haloes by linking particles and substructures when they lie with a linking length of $b\left(V/N\right)^{1/3}$.
We use a value of $b=0.2$, $V$ is the volume of the box and $N$ is the number of particles in the box.
Gravitationally bound subhaloes are identified using \textsc{SubFind}~\citep{Springel2001,Dolag2009}.
The position of a subhalo is determined by the position of its most bound member particle, and the most massive subhalo in a FoF group is identified as the main halo.

\subsection{SIDM model}
\label{sec:methods_model}

In this paper, we explore a DM model where the DM particles are able to self-interact with small cross-sections in addition to gravitational interactions.  The particles scatter only a few times during the formation of the halo, so the DM remains largely collisionless. We employ the \textit{endothermic} model introduced in \citet{O'Neil2023} based on the theory described in Section 3.1 of\citet{Schutz2014}.
The code is the same as used in \citet{Vogelsberger2019}, which modelled a similar two-state model with different cross sections. In this framework, DM particles have two states, and the energy difference between the two states is much less than the ground-state energy. The model is fully described with four parameters: the ground state mass $m_{\chi^1}$, the mass splitting between the ground and excited states $\delta$, the dark force mediator mass $m_\phi$, and the dark coupling constant $\alpha$.
The particles can scatter elastically, where there is no state change, or inelastically, where the particles change states.
There are five ways for particles to scatter in our model:
\begin{enumerate}
    \item \noindent \textit{Elastic scattering of two ground state particles}: \\ 
    $\chi^1 + \chi^1 \to \chi^1 + \chi^1$ \\
    In this reaction, two ground-state particles scatter and remain in the ground state. The total kinetic energy remains the same, but energy can be transferred between the particles and therefore through the halo. Elastic scatterings tend to produce a thermalised core in DM haloes.
    \vspace{0.1cm}
    
    \item \noindent\textit{Inelastic up-scattering of two ground state particles}:\\
    $\chi^1 + \chi^1 \to \chi^2 + \chi^2$\\
    In this reaction, two ground state particles scatter to produce two excited state particles. In this case, kinetic energy from the two particles is transformed to the energy of excited states, which results in ``dark cooling flow'' towards the centre of the halo.
    \vspace{0.1cm}
    
    \item \noindent\textit{Elastic scattering of two excited state particles}: \\
    $\chi^2 + \chi^2 \to \chi^2 + \chi^2$\\
    This reaction is similar to the elastic scattering between ground state particles but occurs between excited state particles. In this model, the cross-section for these two elastic scattering reactions is the same. This model also contributes energy transfer between particles and core creation.
    \vspace{0.1cm}
    
    \item \noindent\textit{Inelastic down-scattering of two excited state particles}:\\ 
    $\chi^2 + \chi^2 \to \chi^1 + \chi^1$\\ 
    In this reaction, two excited state particles scatter to produce two ground state particles. The energy difference between the states is converted to the kinetic energy of DM particle. This can cause the particles to move farther out in the DM halo, or to become unbound entirely.
    \vspace{0.1cm}
    
    \item \noindent\textit{Elastic scattering of a ground and an excited state particle}: \\
    $\chi^1 + \chi^2 \to \chi^1 + \chi^2$\\
    Similar to the other elastic scattering reactions, there is no state change effectively. The cross-section for this reaction is calculated using the analytic cross-sections derived in \citet{Schutz2014} in the Born regime ($\alpha m_{\chi^1}\ll m_\phi$). To account for the heavy mediator in this model, the calculation is supplemented with expressions from \citet{Feng:2009hw}.
\end{enumerate}

The values for the model we adopt in this work are $m_{\chi^1}=2.3\:$MeV, $\delta=0.48\:$eV, $m_\phi=7.3\:$MeV, and $\alpha=0.17$.
These values lead to a significant probability of both up- and down-scattering in addition to elastic scattering between ground and excited state particles.
Elastic scattering between either two ground state or two excited state particles is suppressed though does happen in small amounts.
While there have been some constraints placed on elastic scattering \citep[see][for review]{Adhikari2022}, inelastic models are less constrained.
Additionally, these cross sections are high enough to be on the extreme end of what might be seen, both making the influences clear and testing the limits of the model.

Figure \ref{fig:cross_sections} shows the transfer cross section per unit mass ($\sigma_T/m_\chi$) for each type of scattering.
The transfer cross sections are a function of relative velocity between the two particles. The conversion between halo mass and relative velocity is done by assuming an isotropic Maxwell-Boltzmann velocity distribution of DM particles with one-dimensional velocity dispersion $\sigma_{\rm 1d} \simeq 0.6\,V_{\rm vir}$~\citep{Lokas2001}, where $V_{\rm vir}\equiv \sqrt{GM_{\rm 200,mean}/R_{\rm 200,mean}}$ is the virial velocity of the halo.
The elastic scattering between ground state particles (purple line) coincides with the cross section for elastic scattering between excited state particles (green dashed line) at a value of $0.21\:{\rm cm}^2{\rm g}^{-1}$.
The up-scattering reaction (orange line) is only possible when the relative velocity is high enough such that there is sufficient kinetic energy to convert to the energy difference between the ground and excited states.
At lower velocities, the down-scattering reaction (red line) becomes more likely, scaling with a power law of -1, which can give particles more kinetic energy.
Elastic scattering between a ground and an excited state particle is approximately constant at a value of $15\:{\rm cm}^2{\rm g}^{-1}$.

Within the simulation, individual DM particles are represented with simulation macroscopic particles with a mass as given by the resolution (i.e. $2.2\times10^5\msun$ for zoom particles).
Each DM particle represents an ensemble of fundamental DM particles that are entirely in either the ground or excited state and all of which scatter simultaneously. These particles scatter less frequently than fundamental DM particles such that over the course of the simulation, the expected amount of DM mass experiences each scattering reaction.

We follow the algorithm described in \citet{Schutz2014} and used in \citet{Vogelsberger2019} to implement the SIDM model in the simulations, which we briefly summarise here.
Whether a particle scatters or not, and which scattering occurs, depends on both the scattering transfer cross sections and the local density of dark matter. The local density is determined by the 10 to 38 nearest neighbours smoothed by a cubic spline kernel function around its nearest neighbours.
The cross-section at the relevant velocity between two particles is multiplied by the DM (simulation macroscopic) particle mass and the distance each particle would move in one time step to get the probability of scattering.
The probabilities are appropriately adjusted by the kernel factor as determined by the distance between the two potential scattering particles.
Whether a particle scatters with a nearby particle and which reaction occurs is determined by drawing a random number.

In the event that a scattering reaction occurs, an appropriate second particle is selected to scatter with.
The timesteps are set to be smaller than the expected scattering frequency so that multiple scatters per particle per time step are rare.
In the case that there are multiple scatters in a single time step, the second reaction is rejected and does not take place.
This is because a particle's position and velocity can be updated only once per time step.
Rejected scatters make up less than $1\%$ of the total expected scatters, and further decreasing the time step to avoid these does not impact the results.
We have tested this by running the same initial conditions with larger cross sections and varying the time step and rejection rates under 5 percent do not noticeable alter the particle distribution in the halo.
The particles scatter isotropically, and any changes to the particles' energies are done in the centre of momentum frame of the two particles.

We run a suite of 11 simulations with varying initial excited and ground state fractions between 0\% and 100\%.


\begin{figure}
    \centering
    \includegraphics[width=\linewidth]{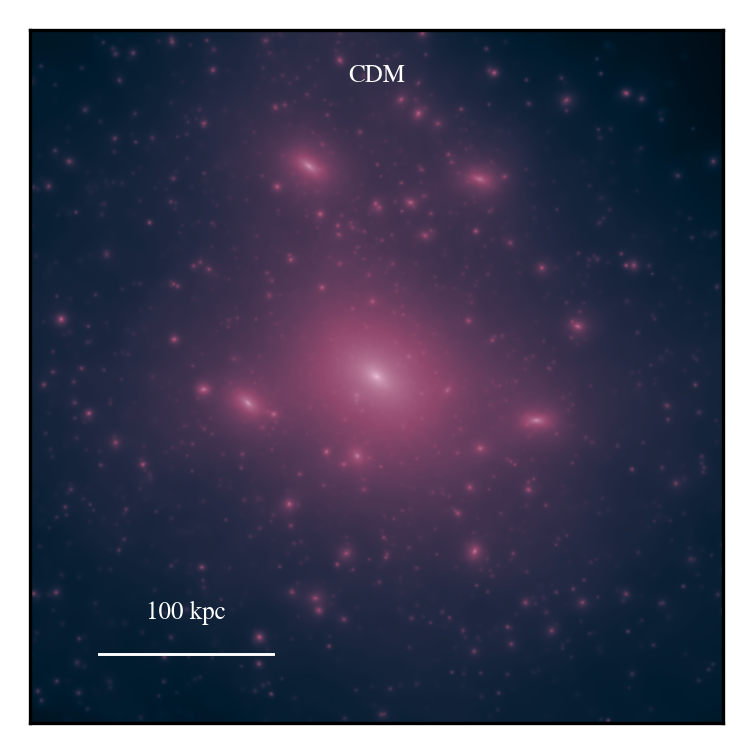}
    \caption{The projected DM density for the main halo of the CDM simulation at $z=0$.  The main halo is surrounded by many subhaloes of various sizes and is slightly oblong.}
    \label{fig:projections_cdm}
\end{figure}

\begin{figure*}
    \centering
    \includegraphics[width=\linewidth]{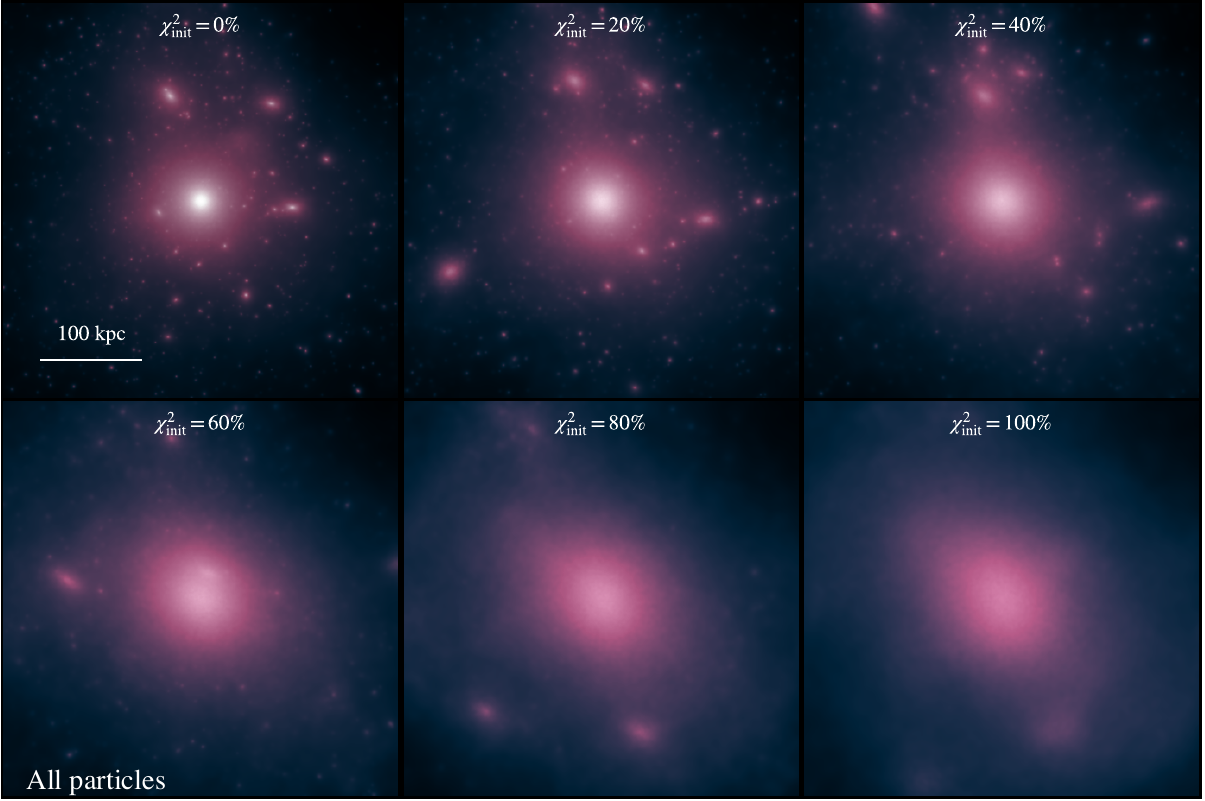}
    \caption{The projected DM density for the main haloes at $z=0$ in the simulations with $\chi^{2}_{\rm init}=$ 0, 20, 40, 60, 80 or 100\% of the particles starting in the excited state. As more particles begin in the excited state, more exothermic reactions occur and the substructure is destroyed.  When more particles begin in the ground state, the endothermic reaction plays an important role in the appearance of the main halo at low redshift, and substructure remains.  We also note that the halos with low excited state initial fractions are more spherical while the higher initial state fraction simulations have a more oblong shape.}
    \label{fig:projections}
\end{figure*}

\begin{figure*}
    \centering
    \includegraphics[width=.49\linewidth]{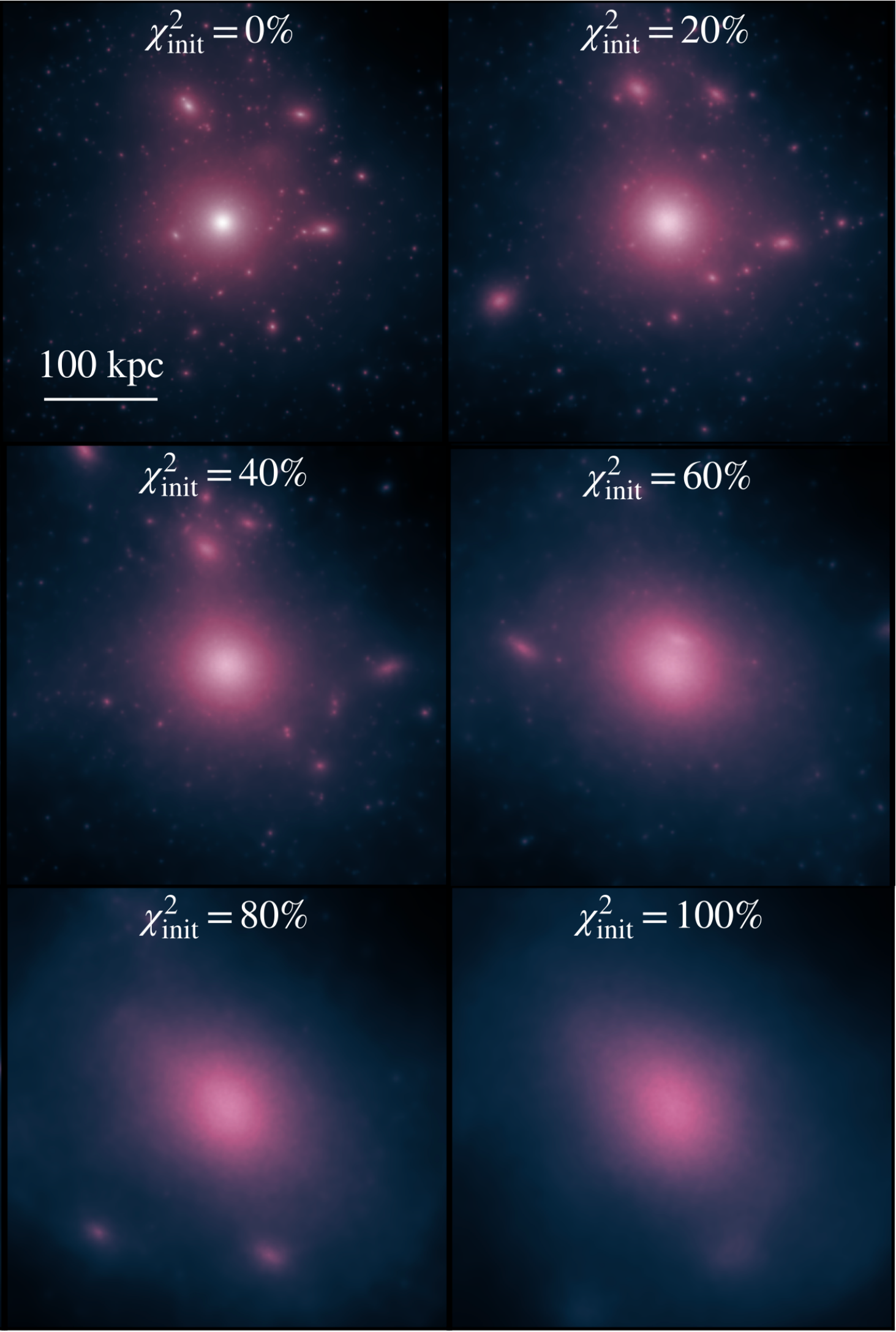}
    \includegraphics[width=.49\linewidth]{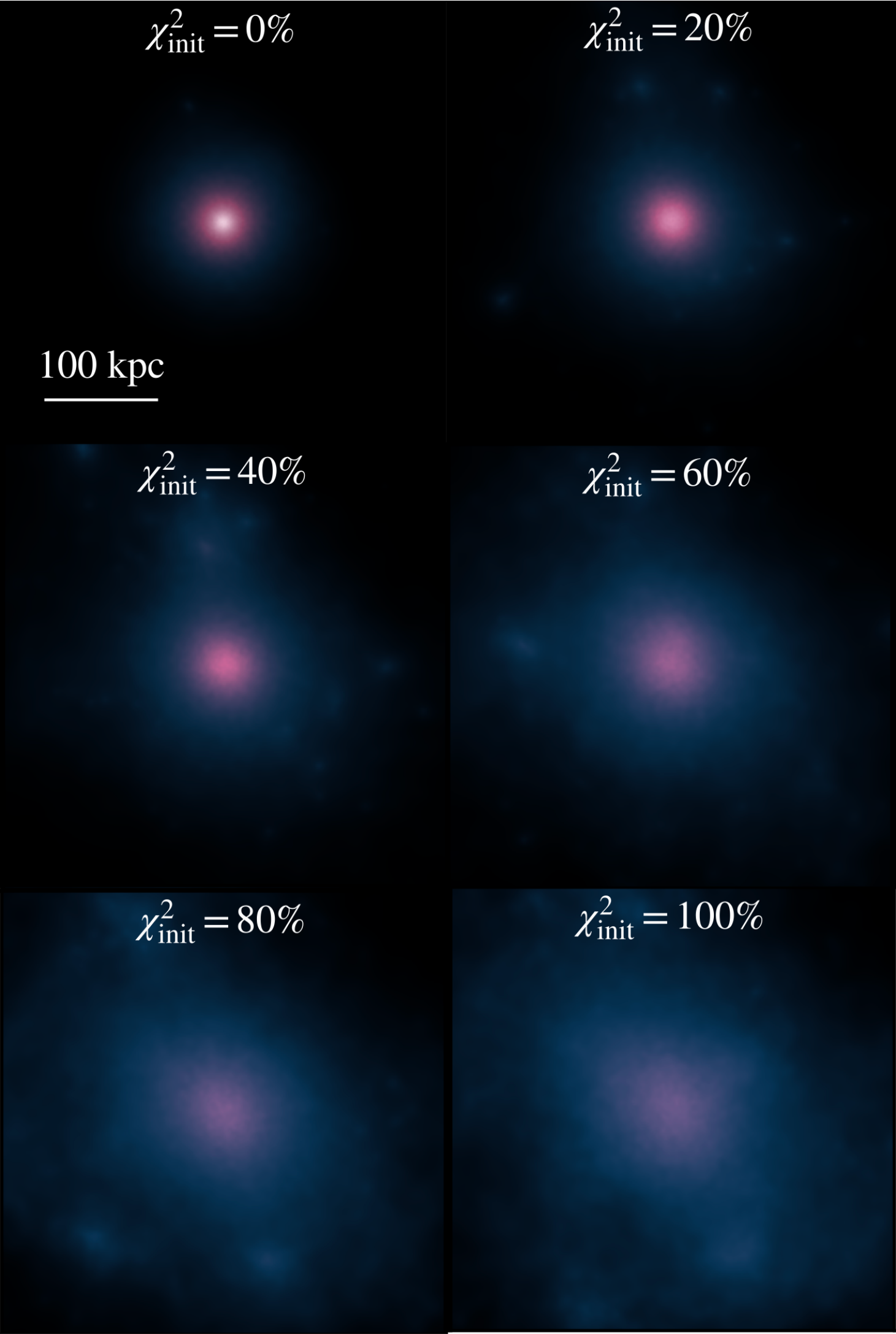}
    \caption{The projected DM density for particles in the ground (left) or excited (right) states in the simulations with 0, 20, 40, 60, 80 or 100\% of the particles starting in the excited state at $z=0$.  Most partices are in the ground state, and the halo does not change much in appearance between the ground state projections and the total particle projections.  The excited state particles occur mostly in the centre of the halo, where the velocities are high enough for up-scattering to occur.  The excited state particles also follow the trend of decreased density as the initial excited state fraction increases.}
    \label{fig:projections_states}
\end{figure*}

\section{Results}
\label{sec:results}

Our suite is made up of 11 simulations with the DM particles initially $0-100\%$, increasing in intervals of $10\%$, in the excited state.
We refer to the initial excited state fraction as $\chi_{\rm init}^2$ and the initial ground state fraction as $\chi_{\rm init}^1$. In Figure \ref{fig:projections_cdm}, we show the DM projection for our fiducial CDM simulation.
In Figure \ref{fig:projections}, we show DM projections for all particles several simulations spanning the range of $\chi_{\rm init}^2$. We show the DM projections for excited and ground state particles separately in Figure \ref{fig:projections_states}.
Each simulation has a similar virial radius, ranging from 370.58 kpc for $\chi^2_\text{init} = 100\%$ to 401.27 kpc for CDM.

\subsection{Effects on halo density profiles}
\label{sec:results_density}

\begin{figure}
    \centering
    \includegraphics[width=\columnwidth]{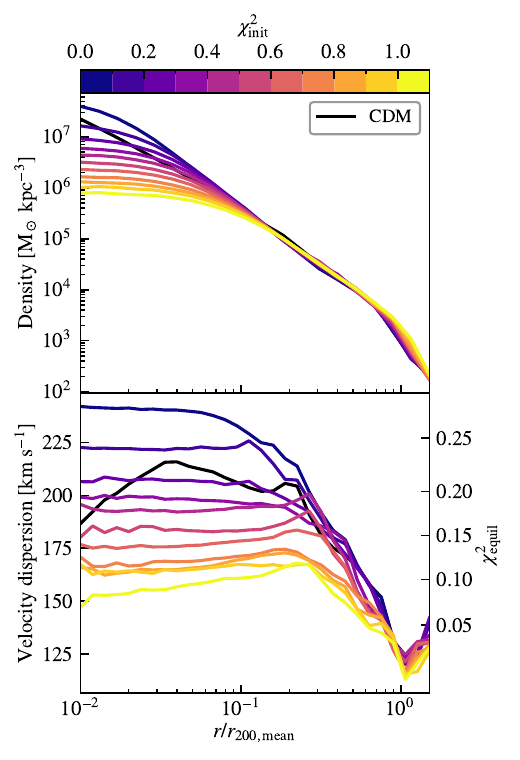}
    \caption{Density profiles and velocity dispersions for the main halo at z=0. Simulations with fewer initially excited particles tend to have denser centers. When all of the particles begin in the ground state, every downscatter must be preceded by an upscatter, so there is more upscattering than downscattering. The right-hand y axis shows the equilibrium excited state fraction, $\chi^2_{\rm{equil}}$, at relevant velocity dispersions, using $\sigma_{\rm{3d}} = \sqrt{3}\sigma_{\rm{1d}}$.}
    \label{fig:density_vdisp}
\end{figure}

The mass distribution within the haloes varies significantly. In Figure~\ref{fig:density_vdisp}, we plot density profiles for the main halo at $z=0$ in the top panel. The SIDM model gives a cored inner density profile for all initial excited fractions, compared to the cuspy profile of CDM. This is similar to the effects of classical elastic SIDM models previously explored in literature~\citep[e.g.][]{Zavala2013, Rocha2013, Elbert2015} and is due to thermalisation of DM cores from heat conduction.

The two-state configuration of the SIDM model studied in this paper enables additional channels of heating or cooling in the dark sector. As shown in Figure~\ref{fig:density_vdisp}, when $\chi^2_\text{init}$ is low, we find that the central density of the simulated Milky Way-mass halo is enhanced compared to CDM. This is similar to the gravothermal core-collapse studied in lower halo mass regimes and often in SIDM models with strong velocity-dependence of cross sections~\citep[e.g.][]{Nadler2020,Sameie2020, Correa2022,Zeng2022}. The up-scatterings of DM particles dissipate away their kinetic energy and can accelerate the gravothermal collapse of the halo~\citep[e.g.][]{Essig2019}. This is a common phenomenon found in previous explorations of dissipative SIDM~\citep[e.g.][]{Essig2019, Shen2021, Roy2023}. For example, 
\citet{Shen2021} simulated a DM model that dissipates a constant fraction of a particle's kinetic energy during an interaction.
They ran a suite of zoom-in simulations of dwarf to Milky Way-mass galaxies with different cross sections in addition to a fiducial CDM model, including the baryonic physics described by the FIRE-2 model~\citep{Hopkins2018}. They found that the halo central density increases with cross section and effectively the energy dissipation rate of DM. Similar to the up-scattering in our work, the dissipation causes a particle to lose velocity and fall inwards, thus increasing the central density.

\begin{figure}
    \centering
    \includegraphics[width=\columnwidth]{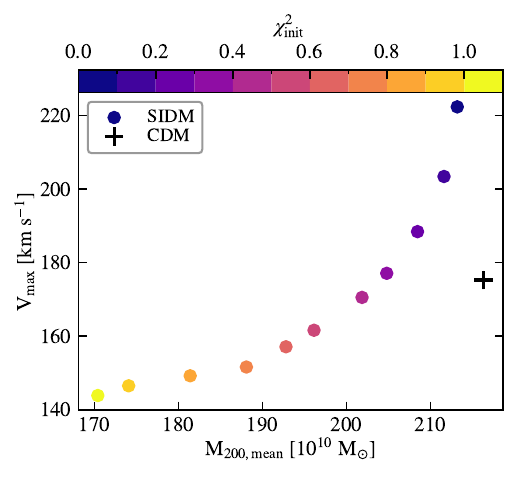}
    \caption{M$_{200,\rm mean}$ and V$_{\rm max}$ for the main halo in each simulation. M$_{200,\rm mean}$ decreases with $\chi^2_{\rm init}$ due to the kinetic energy boost from down-scattering, which can push particles out of the halo. V$_{\rm max}$ also decreases with $\chi^2_{\rm init}$, reflecting trends in both halo mass and concentration. Haloes with low $\chi^2_{\rm init}$ experience more up-scattering, driving particles inwards.}
    \label{fig:v_max-m_200}
\end{figure}

However, increasing $\chi^2_\text{init}$ leads to a lower core density and higher outer density. The half mass radius $r_{1/2}$ increases significantly with $\chi^2_\text{init}$, while the virial radius $r_{200,\text{mean}}$ decreases slightly. As shown in the top panel of Figure~\ref{fig:density_vdisp}, simulations with higher initial excited state fraction $\chi^2_\textrm{init}$ have lower inner densities, varying overall by two orders of magnitude. Exothermic reactions from down-scatterings dominate in these cases and decrease the central density, as previously explored in $N$-body zoom-in simulations~\citep{Vogelsberger2019}. An initially larger fraction of excited particles eventually provides heating of DM particles through downscattering, pushing more particles away from the halo center. This also leads to higher densities at $R_{200}$ for simulations with higher $\chi^2_\textrm{init}$, although the difference is less dramatic. The velocity boost also causes virial mass ($M_{200}$) to decrease with $\chi^2_\textrm{init}$ as more particles are driven above escape velocity as shown in Figure \ref{fig:v_max-m_200}.

The gradients of the velocity dispersions in the SIDM models vanish at lower radii while the CDM peaks at about $10$\:kpc.
Because heat can transfer between DM particles in a self-interacting model, the velocity dispersion smooths out,
The equilibrium velocity dispersion decreases with $\chi_{\rm init}^2$, and it is higher than the CDM case for low $\chi_{\rm init}^2$.
This implies that there is heat transferred out of the halo when particles start in the excited state.
Initial down-scattering pushes particles outwards and out of the halo, leaving the halo at a slightly lower mass and with less energy.
In the low $\chi^2_{\rm init}$ simulations, on the other hand, up-scattering earlier in the halo's evolution creates a deep gravitational well that binds particles within the halo.
The mass of the $\chi_{\rm init}^2=0\%$ is $2.16\times10^{12}\:\msun$ and $\chi_{\rm init}^2=100\%$ is $1.70\:\msun$, and the mass decreases monotonically between the two.

The outwards transfer of energy from down-scattering in the high $\chi_{\rm init}^2$ also increases the core size while the earlier up-scattering in low $\chi_{\rm init}^2$ increases the core density value.
When a particle down-scatters, it increases its kinetic energy by the energy difference between the ground and excited state $\delta=10\:$keV, so the radius in the gravitational well increases by
\begin{equation}
    \begin{aligned}
        \delta &= \frac{GM_{\rm enc}m_{\chi^1}}{R} - \frac{GM_{\rm enc}m_{\chi^1}}{R+\Delta r} \\
        &\approx \frac{GM_{\rm enc}m_{\chi^1}}{R}\frac{\Delta r}{R} \\
        \Delta r &= \frac{\delta R^2}{GM_{\rm enc}m_{\chi^1}}
    \end{aligned}
\end{equation}
which comes to $\Delta r \approx 0.4\:$kpc.

In Figure~\ref{fig:v_max-m_200} we show M$_{\rm 200,mean}$ and V$_{\rm max}$ for the main halo in each simulation. $V_{\rm max}$ is the maximum circular velocity of the halo. As a result of dominant upscatterings, SIDM models with low $\chi_{\rm init}^2$ values produce more concentrated DM distributions and thus give larger $V_{\rm max}$ values. On the contrary, models with large $\chi_{\rm init}^2$ values feature substantial downscatterings of DM particles that create cored density profiles. The cores not only decrease the $V_{\rm max}$ values in these models but also lead to lower halo virial radius and masses (given the same spherical overdensity threshold).

\subsection{Excited state fraction}
\label{sec:results_excited_fraction}

\begin{figure}
    \centering
    \includegraphics[width=\columnwidth]{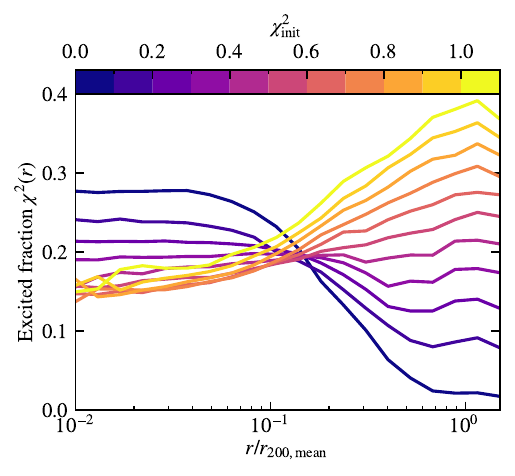}
    \caption{The fraction of main halo particles in the excited state as a function of radius at $z=0$. In the outer region $(r \gtrsim 100 \text{ kpc})$, $\chi^2$ increases with $\chi^2_\text{init}$. This is due to the lower density at these radii, which decreases the impact of scattering. At smaller radii, $\chi^2$ is not monotonic over $\chi^2_\text{init}$, but simulations with higher $\chi^2_\text{init}$ counterintuitively tend to have lower $\chi^2(z=0)$. This is likely related to the lower inner velocity dispersion in these simulations, which causes down-scattering to dominate.}
    \label{fig:states_r}
\end{figure}

Figure~\ref{fig:states_r} shows the fraction of particles in the excited state at redshift 0 $\left(\chi^2(z=0)\right)$ as a function of radius $r$. At small radii, $\chi^2(z=0)$ shows the opposite trend with respect to $\chi^2_\text{init}$, i.e. models with higher initial excited state fraction end up with lower excited state fraction in halo centers at $z=0$. This is likely due to the different central velocity dispersions of DM particles reached in different models, which results in different equilibrium states. To quantitatively understand this phenomenon, we can approximate the local equilibrium excited state fraction by equating the upscattering and downscattering rates. The expected time for one ground-state particle to upscatter once is
\begin{equation}
    t_{\rm{up}} = \dfrac{1}{\langle\rho_{\rm{gr}}v_{\rm{rel}}\frac{\sigma_{\rm{up}}}{m}\rangle},
\end{equation}
where $\langle ... \rangle$ denotes the thermal average, $\rho_{\rm gr}$ ($\rho_{\rm ex}$) denotes the density of ground (excited) state DM particles, and $\sigma_{\rm up}$ is the upscattering transfer cross section. For simplicity, we assume that the velocities of DM particles locally obey the Maxwell-Boltzmann distribution and we obtain
\begin{equation}
    \langle X \rangle = \dfrac{1}{2\sqrt{\pi} \sigma_{\rm 1d}^3}\int_0^{\infty}{\rm d}v_{\rm rel} v_{\rm rel}^{2} e^{-v_{\rm rel}^2/4\sigma_{\rm 1d}^2} X,
    \label{eq:thermal_average}
\end{equation}
where $\sigma_{\rm 1d}$ is the local one-dimensional velocity dispersion of DM. Therefore, the expected up-scattering rate in a unit volume can be expressed as
\begin{equation}
    \Gamma_{\rm{up}} = n_{\rm gr}\, \langle\rho_{\rm{gr}}v_{\rm{rel}}\frac{\sigma_{\rm{up}}}{m}\rangle =  \dfrac{(\chi^1)^2}{m}\, \rho^2\langle v_{\rm{rel}}\frac{\sigma_{\rm{up}}}{m}\rangle.
\end{equation}
And the same rate for downscattering is
\begin{equation}
     \Gamma_{\rm{down}} = \dfrac{(\chi^2)^2}{m}\, \rho^2\langle v_{\rm{rel}}\frac{\sigma_{\rm{down}}}{m}\rangle
\end{equation}

A local equilibrium occurs when $\Gamma_{\rm{up}} = \Gamma_{\rm{down}}$ and we obtain the equilibrium excited state fraction as
\begin{align}
    \left(\chi^1/\chi^2\right)_{\rm eq.}^2 &= \langle v_{\rm{rel}}\frac{\sigma_{\rm{down}}}{m}\rangle / \langle v_{\rm{rel}}\frac{\sigma_{\rm{up}}}{m}\rangle \nonumber \\
    &= \frac{\frac{1}{\sigma^3_{\rm{1d,ex}}}\int_0^{\infty}\rm{d}v_{\rm{rel}}v^3_{\rm{rel}}e^{-v^2_{
\rm{rel}}/4\sigma^2_{\rm{1d,ex}}}\frac{\sigma_{\rm{down}}}{m}}{\frac{1}{\sigma^3_{\rm{1d,gr}}}\int_0^{\infty}\rm{d}v_{\rm{rel}}v^3_{\rm{rel}}e^{-v^2_{
\rm{rel}}/4\sigma^2_{\rm{1d,gr}}}\frac{\sigma_{\rm{up}}}{m}}
\label{eq:chi_eq}
\end{align}
where $\sigma_{\rm{1d,gr}}$ and $\sigma_{\rm{1d,ex}}$ are one-dimensional velocity dispersions of ground and excited state particles, respectively. For simplicity, we assume that both of them are the same as the total DM velocity dispersion $\sigma_{\rm 1d}$. On the right $y$-axis of Figure~\ref{fig:density_vdisp}, we show the predicted equilibrium $\chi^2$ as a function of $\sigma_{\rm{1d}}$, assuming that excited and ground state particles have the same velocity dispersion. At asymptotically high velocities where the probability for up- and down-scattering are consistently similar, we should get $50\%$ of the particles in the excited state.
For the velocity dispersion of our simulated haloes, which is $\sigma_{\rm 3D} \sim 200-300 \kms$ so the one-dimensional velocity dispersion is $\sigma_{\rm 1d} \sim 100-150 \kms$, corresponding to an excited state mass ratio of $0.1-0.3$, which is consistent with the excited state fractions at lower radii in Figure~\ref{fig:states_r}.

On the other hand, the excited state fraction at the outskirt $(r \gtrsim 100\text{ kpc})$ of the halo in Figure~\ref{fig:states_r} follows a qualitatively different trend. $\chi^2(z=0)$ increases monotonically with $\chi^2_\text{init}$ and a crossing point of the $\chi^2(z=0)$ exists around $0.1-0.2\,R_{\rm 200,mean}$. Since the density and velocity dispersion of DM at large radii are low enough such that the up-scattering channel is completely suppressed (see Figure~\ref{fig:cross_sections}), the equilibrium solution cannot be realized. The excited state fraction will simply move monotonically downwards with time due to downscattering, resulting in the trend seen at large radii. The cosmological evolution of excited particle number density when the upscattering channel is turned off would be
\begin{equation}
    \dfrac{{\rm d}n_{\rm ex}(t)}{{\rm d}t} + 3\,H(t)\,n_{\rm ex}(t)= - 2\,n_{\rm ex}(t)^{2}\,m_{\chi^1}\,\langle v_{\rm{rel}}\frac{\sigma_{\rm{down}}}{m_{\chi^1}}\rangle,
    \label{eq:nex_evol}
\end{equation}
with $n_{\rm ex}(z_{\rm i}) = \chi^2\,n_{\rm dm}(z_{\rm i})$ where $n_{\rm dm}$ is the number density of all DM particles. We take $z_{\rm i}$ to be the starting redshift $z_{\rm i}=127$ of the simulations. As shown in Figure~\ref{fig:cross_sections}, in the low-velocity end, the product of particle relative velocities and self-interaction cross-sections tend to have a stabilized value around $6\times 10^{3}\,{\rm cm}^{2}\,{\rm g}^{-1}\,\kms$. Integrating Equation~\ref{eq:nex_evol} and comparing to the $n_{\rm dm}(z=0)$, we obtain $\chi^2 \sim 0.1,\, 0.3,\, 0.4$ of excited particles at $z=0$ in models with $\chi^2_{\rm init}=0.1,\, 0.5,\,0.9$, respectively. This agrees with the excited fraction we found at the outskirts of the simulated haloes. Such a picture also predicts that the excited state fraction in the low-density region of the Universe decreases much faster at earlier times due to higher particle number densities and stabilizes to $\lesssim 2\%$ level from $z=6$ to $z=0$, which is consistent with our findings in Section~\ref{sec:results_redshift}.

\subsection{Halo shapes}
\label{sec:results_shape}

Here we discuss how the morphology of the halo as a function of $\chi_{\rm init}^2$.
As seen in Figure \ref{fig:projections}, there is a noticeable difference in the shapes of the haloes as the initial excited state fraction varies.
The haloes with a lower $\chi_{\rm init}^2$ are more concentrated and spherical than those with higher $\chi_{\rm init}^2$.

\begin{figure}
    \centering
    \includegraphics[width=\columnwidth]{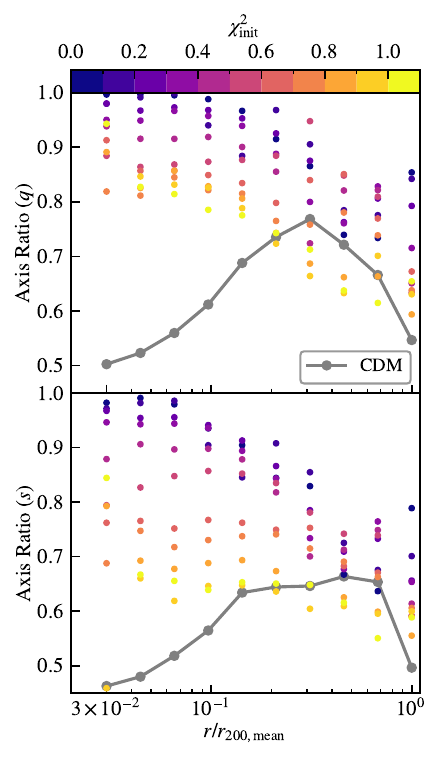}
    \caption{We fit an ellipse to each halo at several radii and find the ratio of the middle axis ($q$, top panel) and minor axis ($s$, bottom panel) to the major axis. A ratio of $s=1$ is spherical, and a ratio of $s=0$ is flat.  The grey points show the CDM run.  Each column of points is the axis ratio at that radius for each simulation halo.  We see that low $\chi_{\rm init}^2$ haloes tend to be spherical due to early up-scattering that compresses the halo.  High $\chi_{\rm init}^2$ haloes maintain a more oblong shape since down-scattering keeps the halo enlarged.}
    \label{fig:axis-ratios-all}
\end{figure}

Figure \ref{fig:axis-ratios-all} demonstrates this more quantitatively.
For each halo, we fit an ellipse at different radii from 0.03 $R_{\rm 200,mean}$ to $R_{\rm 200,mean}$ and divide the middle and minor axes by the major axis to measure how spherical it is.
We follow the method used in \citet{Chua2021}, which iteratively fits an ellipsoidal shell to the halo while keeping the semi-major axis constant. First, we fix the semi-major axis, $R$.
At each step, the bin is set to be within 0.1 dex of the current ellipse.

We begin with a spherical bin, defined by $A_{ij} = \delta_{ij}$. We calculate the shape tensor of the particles inside the bin, defined as
\begin{equation}
S_{ij} = \frac{\sum_{k=1}^Nm^kx^k_ix^k_j}{\sum_{k=1}^Nm^k}
\end{equation}
Next, we obtain the eigenvectors $v_1,v_2,v_3$ with eigenvalues $\lambda_1,\lambda_2,\lambda_3$ respectively ($\lambda_1 \geq \lambda_2 \geq \lambda_3$). These determine the intermediate axis ratios $q \equiv \sqrt{\lambda_2/\lambda_1}$ and $s \equiv \sqrt{\lambda_3/\lambda_1}$.
We also rotate the particles to a coordinate system with axes aligning with the eigenvectors.
Finally, we generate a new bin with the new eigenvalues determining $q$ and $s$. The iteration terminates when $q$ and $s$ change by less than $0.1\%$ between steps. For more details, see \citet{2011ApJS..197...30Z}. 
A ratio of $s=1$ is perfectly spherical, and smaller ratios are increasingly oblong.
At each radius, we plot $q$ and $s$ for each $\chi_{\rm init}^2$ in different colours.

Across all radii, haloes with a lower $\chi_{\rm init}^2$ are more spherical than haloes with a higher $\chi_{\rm init}^2$.
This is consistent with previous results in \citet{O'Neil2023}, where the endothermic model resulted in a more spherical halo than the exothermic model.
Here, higher $\chi_{\rm init}^2$ haloes are dominated by exothermic reactions since the down-scattering reaction is the only available reaction when all particles are in the excited state.
In elastic scattering, transfer cross sections of approximately $\sigma_T/m_\chi=0.1\:{\rm cm}^2{\rm g}^{-1}$ drive spherical centres in the cores of galaxy clusters \citep[e.g.][]{Muralda-Escude2002,Peter2013,Brinckmann2018,Shen2022-sidm}.
Here, we show that significant endothermic and elastic scattering produces similar results, but the exothermic scattering mitigates this, allowing larger elastic transfer cross sections while maintaining a certain amount of asymmetry.
To fully understand the constraints in galaxy clusters, a dedicated set of simulations is needed to accurately predict how this model would behave in such environments.

Haloes do not accrete symmetrically, so it is expected that there is a certain amount of asymmetry as seen in the high $\chi_{\rm init}^2$ simulations.
This asymmetry is wiped away by the endothermic reactions.
Endothermic reactions cause particles to fall deeper into the potential well by reducing their kinetic energy during the reaction.
This compresses the inner regions of the halo and causes it to become more spherical, especially when a large amount of particles are up-scattered.
This is consistent with findings that low-concentration halos are generally less symmetrical than high-concentration halos \citep{Hashimoto2007}
Even when these particles are later down-scattered and kicked further out in the halo, they cannot regain the asymmetrical orbits they previously may have had and maintain the spherical shape of the halo.

\citet{Chua2019} studied halo shapes in Illustris and Illustris-Dark using methods similar to ours and focused on the difference between results with and without radiation.
In CDM haloes in $N$-body simulations, they found, similar to our results, that the halo is least spherical in the centre, with a median axis ratio of 0.6.
They also found that radiation increases how spherical the halo is.
We find that SIDM, both up- and down-scattering dominant, increase the how spherical a halo is.
However, unlike radiation, we find that SIDM alters the shape of the axis ratio as a function of radius.

In particular, our SIDM haloes become much more spherical in the centre compared to the CDM.
This is likely due to the up-scattering reaction in our SIDM model.
Even in the high $\chi_{\rm init}^2$ haloes, particles that have up-scattered are more likely to be found closer to the centre of the halo.
Up-scattering is also the reaction that increases the axis ratio, which leads to our trend of increasing axis ratio towards the centre of the haloes.

\begin{figure}
    \centering
    \includegraphics[width=\linewidth]{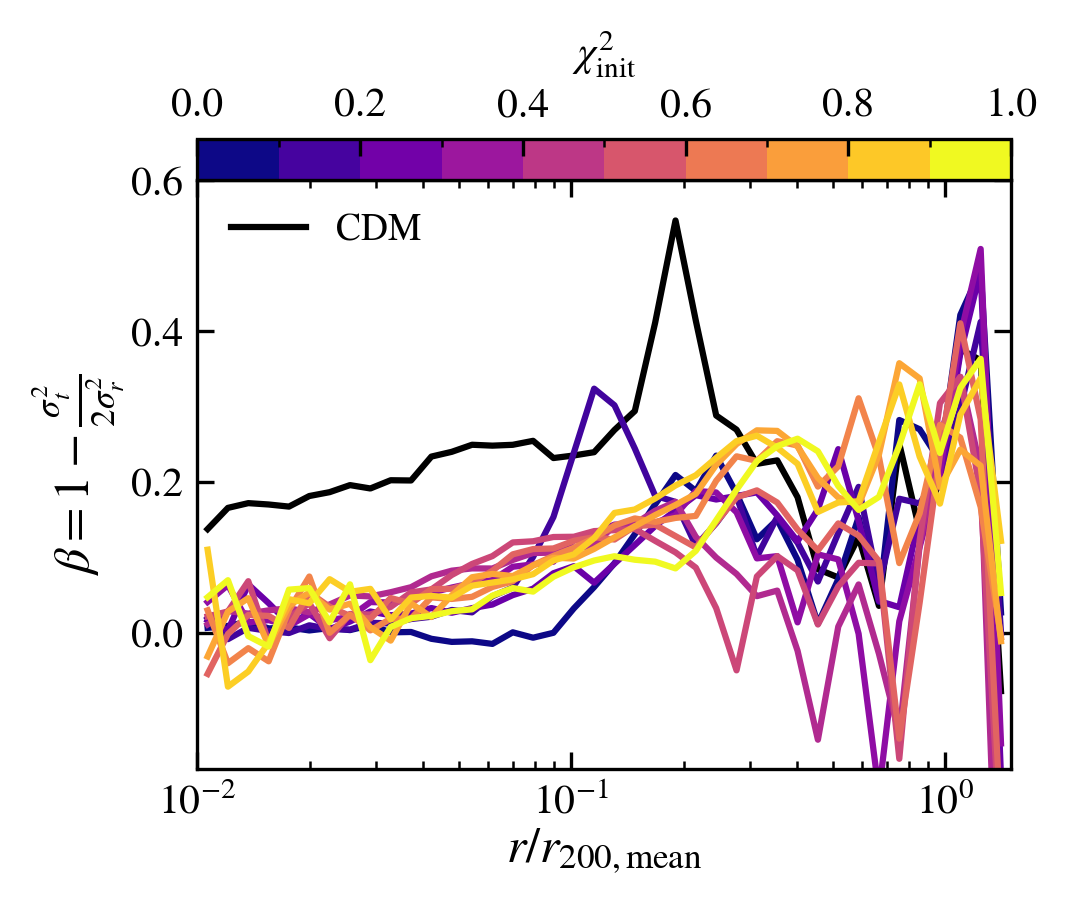}
    \caption{The anisotropy parameter as a function of radius from the halo centre.  A larger $\beta$ indicates a more radial orbit.  The SIDM haloes have a lower $\beta$ than the CDM halo due to the energy transfer between particles, which causes the orbits to be more isotropic.  In particular, the SIDM anisotropy curves reach similar values at low radii, where they have similar cored slopes.}
    \label{fig:anisotropy}
\end{figure}

\subsection{Kinematic properties}
\label{sec:results_kinematic}

The bottom panel of Figure~\ref{fig:density_vdisp} shows the velocity dispersion for all main halo particles as a function of radius. The velocity dispersion indicates the relative velocities between nearby particles, which determines the frequency of each type of scattering. 
Simulations with higher $\chi^2_\text{init}$ tend to have lower velocity dispersions in the inner regions, which corresponds to a higher frequency of downscattering and a lower frequency of upscattering. 

In Figure \ref{fig:anisotropy}, we show the anisotropy parameter
\begin{equation}
    \beta \equiv 1 - \frac{\sigma_{\rm t}^2}{2\sigma_{\rm r}^2}
\end{equation}
of the particle velocities in each simulation's main halo.
To calculate the anisotropy, we first calculate the radial and two components of angular velocity.
We then take the variance in the radial component to calculate $\sigma_{\rm r}$ and then sum the variance of the two angular components to get the tangential variance $\sigma_{\rm t}$.

The CDM curve follows the anistropy as a function of radius that is typically found \citep[e.g.][]{Navarro2010}.
Once self-interactions are introduced, the orbits become much more radial.
As inelastic reactions occur, the total velocity changes and the particles are scattered isotropically.
However, because of the change in kinetic energy, the particles must also move either inwards or outwards in the gravitational potential.
This causes the orbits to become more radial compared to any tangential change the particles experience.

\citet{Hansen2006} noted that there is a strong correlation between $\beta$ and the slope of the density profile, and \citep{Navarro2010} found that this holds primarily for the inner regions of the halo and not the outer regions.
In particular, they find that $\beta$ and $\frac{{\rm d}\log\rho}{{\rm d}\log r}$ are negatively correlated, with higher slopes, i.e. more cored, corresponding to lower $\beta$.
In Figure \ref{fig:anisotropy}, we find that the SIDM models all have significantly lower $\beta$ compared to the CDM model, corresponding to a higher slope that is closer to zero.
Although the densities of the SIDM models vary, they all flatten to a core in the centre and their anisotropy values are correspondingly similar at low radii.
At larger radii, the SIDM anistropy peaks as the CDM does, though the peak is much wider.
This is likely due to the energy transfer that is allowed between SIDM particles, which causes the velocities to be more uniform.
The anisotropy is also generally lower at large radii for lower $\chi_{\rm init}^2$.

\subsection{Effects on substructures}
\label{sec:results_substructure}

\begin{figure}
    \centering
    \includegraphics[width=\columnwidth]{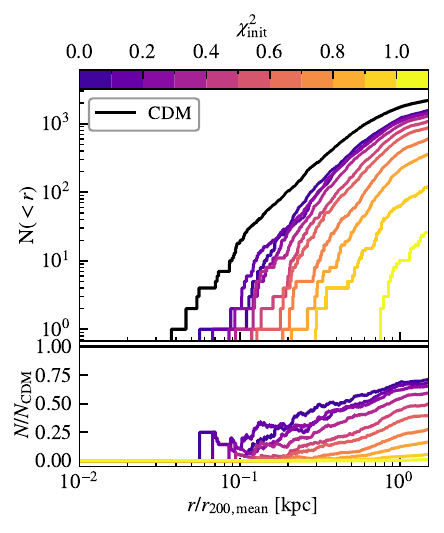}
    \caption{{\it Top}: The number of subhaloes with mass > $10^7 M_\odot$ within a radius $r$ of the main halo at $z=0$.  As $\chi_{\rm init}^2$ increases, the number of haloes  decreases, although the shape of the curve remains the same.  This is due to down-scattering that gives a velocity kick to particles and makes it more difficult to keep them within a gravitational well.  Even at $\chi_{\rm init}^2 = 0$, we see a decrease in substructure due to tidal forces.
    {\it Bottom}: The fractional difference in the subhalo number profile compared to CDM.  The $\chi_{\rm init}^2=0$ halo has approximately half as many subhaloes at each radius as the CDM halo, and this fraction decreases as $\chi_{\rm init}^2$ increases.}
   \label{fig:subhalo_counts_radius}
\end{figure}

Figure~\ref{fig:subhalo_counts_radius} shows the cumulative number of subhaloes with mass $M_{\rm sub} > 10^7 \msun$ within a radius $r$ to the main halo centre. In the bottom panel, we show the same quantity divided by that of the CDM simulation. In models with increasing $\chi_{\rm init}^2$, there are significantly less substructures. The down-scatterings increase the kinematic energy of the subhalo and make it more difficult for subhalo with small gravitational wells to stay bound. On the other hand, up-scattering decreases the velocity but the velocity dispersion in satellite haloes is generally not high enough to significantly alter the substructure in up-scattering dominant models. The simulations with higher $\chi^2_{\rm init}$ therefore have a much more altered substructure compared to CDM compared to simulations with a higher $\chi^1_{\rm init}$. 

However, there is still a decrease by a factor of 2 in even the $\chi_{\rm init}^2=0$ case since the main halo increases in central density, which increases the tidal forces and strips the subhaloes.
The tidal radius $r_t$ of the subhalo decreases as the enclosed mass of the main halo increases
\begin{equation}
    r_t = r \left(\frac{m_{\rm sub}}{2M_{\rm main}\left(<r\right)}\right)^{\frac{1}{3}}
    \label{eq:tidal_radius}
\end{equation}
where $m_{\rm sub}$ is the mass of the subhalo, $M_{\rm main}$ is the mass of the main halo and $r$ is the distance of the subhalo from the centre of the main halo \citep{King1962,Pace2022}.
Thus, an increased central density of 10 will result in approximately a factor of 2 decrease in a satellite's tidal radius.

Another important effect is the SIDM ram pressure stripping due to self-interactions between satellite and host halo DM particles~\citep[e.g.][]{Kummer2018, Nadler2020, Jiang2021}, also known as SIDM evaporation~\citep{Zeng2022}. Thermalized cores generated by self-interactions can make satellites more prone to this. The relative importance of ram pressure stripping can vary when the cross-section has velocity dependence~\citep[e.g.][]{Banerjee2020,Nadler2020,Zeng2022}. In our model, the cross-section of the dominant down-scattering process decreases by orders of magnitude with increasing velocities until reaching $\sim 1000 \kms$, which is larger by a factor of few than the maximum circular velocities of the simulated Milky Way-mass halos. The heat generated from scatterings of satellite particles dominates the heat injected from scattering host halo particles. 

\begin{figure}
    \centering
    \includegraphics[width=\columnwidth]{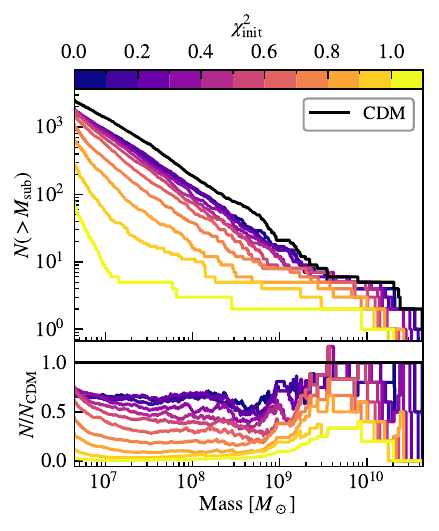}
    \caption{{\it Top}: the number of subhaloes above a given mass in the main halo at $z=0$.  The number of subhaloes decreases with $\chi_{\rm init}^2$ due to down-scattering kicking particles out of the gravitational wells.  At high masses, this effect is less pronounced since the gravitational well is deeper and it is more difficult to escape it.  At low $\chi_{\rm init}^2$, tidal stripping plays a larger role in decreasing the subhalo number than scattering effects.
    {\it Bottom}: The fractional difference in subhalo number compared to the CDM halo. The $\chi_{\rm init}^2=0$ halo has approximately half as many subhaloes at each mass as the CDM halo, and this fraction decreases as $\chi_{\rm init}^2$ increases.}
   \label{fig:subhalo_mass_function}
\end{figure}

Figure~\ref{fig:subhalo_mass_function} shows the cumulative subhalo mass function for the main halo in each solution (top) and the same quantity divided by that of CDM (bottom). We find that simulations with higher $\chi^2_\textrm{init}$ tend to have fewer subhaloes at all mass scales. We can do a simple estimate of the typical subhalo mass where the heat generated from down-scattering exceeds the gravitational binding energy. For a halo with an NFW profile, the gravitational binding energy is \citep{Mo1998}
\begin{equation}
    E_{\rm b} = \gamma(c_h) \dfrac{G M_{200}^2}{R_{200}} = \gamma(c_h)\,G\,\left( \dfrac{4\pi}{3}\,200\,\rho_{\rm mean}\right)^{1/3}\,M^{5/3},
\end{equation}
where $c_{\rm h}$ is the concentration of the subhalo, and
\begin{equation}
    \gamma(c_h) \equiv \dfrac{c_h}{2} \dfrac{1 - 1/(1+c_h)^2 - 2 \ln{(1+c_h)/(1+c_h)}}{\left[c_h/(1+c_h)-\ln(1+c_h)\right]^2} \;.
\end{equation}
Meanwhile, the total energy of excited state particles in the subhalo $E_{\rm ex}$ is
\begin{equation}
    E_{\rm ex} = \delta (\chi^{2}) \dfrac{M_{\rm sub}}{m_{\chi}}.
\end{equation}
This represents the maximum possible heating energy to the subhalo through exothermic self-interactions. Therefore, the smallest substructure we expect to survive this heat injection is
\begin{align}
    M_{\rm sub}^{\rm lim} & = \left( \dfrac{(\delta/m_{\chi})\,(\chi_{\rm init}^2)\,c^2}{\gamma(c_h)\,G} \right)^{3/2} \, \left( \dfrac{4\pi}{3}\,200\,\rho_{\rm mean} \right)^{-1/2} \nonumber \\
    & \simeq 5 \times 10^{10} \msun \left( \dfrac{\delta/m_{\chi}}{2.1\times 10^{-7}}\right)^{3/2}\, \left( \dfrac{\chi^2_{\rm init}}{0.2} \right)^{3/2}\, \left( \dfrac{\gamma(c_h)}{2} \right)^{-3/2}.
\end{align}
This is generally above the subhalo masses in simulations and, therefore, the numbers of subhaloes at all mass scales display significant suppression. We note that this estimation assumes all of the energy in excited states is transformed into the kinetic energy of DM particles and should be interpreted as a limiting case. It also assumes that the down-scattering of DM particles within the satellite is the major channel of heat generation, which may not hold for other types of SIDM models or more massive host halos.

In observations of satellite galaxies around the Milky Way and Andromeda \citep[e.g.][]{McConnachie2012,Lovell2023}, galaxies are measured with dynamical masses well below $10^7\:\msun$.
In addition, \citet{McConnachie2012} measures over 60 galaxies with masses larger than $10^7\:\msun$.
Tentatively speaking, these observations exclude our model with $\chi_{\rm init}^2\gtrsim80\%$.

\subsection{Redshift evolution}
\label{sec:results_redshift}

\begin{figure}
    \centering
    \includegraphics[width=\columnwidth]{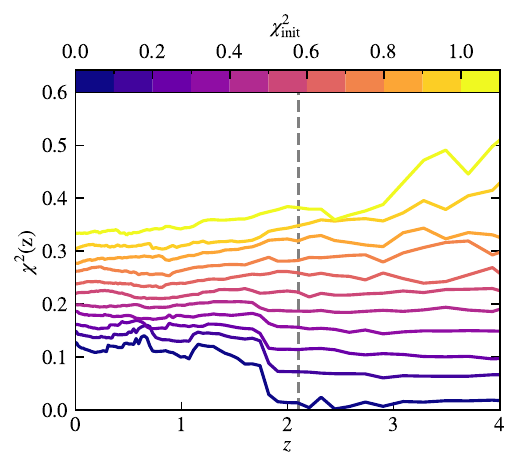}
    \caption{The fraction of main halo particles that are excited by redshift, in the low-resolution simulations. By $z = 0$, each line is approximately flat, indicating that the densest regions of the halo have reached equilibrium. At $z \approx 2$, simulations with low $\chi^2_\textrm{init}$ have a sharp increase in $\chi^2$, corresponding to a halo merger (grey dashed line) that increases the velocity dispersion and the rate of upscattering.}
   \label{fig:z_states}
\end{figure}

Figure~\ref{fig:z_states} shows the fraction of main halo particles in the excited state by redshift, with color denoting $\chi^2_\textrm{init}$.
The simulations converge to more similar values as time goes on, with high $\chi_{\rm init}^2$ haloes decreasing in $\chi^2$ and low $\chi_{\rm init}^2$ haloes increasing in $\chi^2$.
The lines do not cross each other, indicating a smooth transition between dominant exothermic and endothermic reactions.
The final equilibrium state for the haloes is set by the velocity dispersion and the density of particles as described in Equation \ref{eq:nex_evol}.
Thus, although down-scattering is more likely than up-scattering for the higher $\chi_{\rm init}^2$ haloes with lower velocity dispersions, and lower density, particularly on the outskirts of the haloes, prevents significant scattering of either kind.
Therefore, the equilibrium $\chi^2$ remains higher for haloes with higher $\chi_{\rm init}^2$.

Simulations with lower $\chi^2_\textrm{init}$ see a jump in up-scattering near $z = 2$, corresponding to a major halo merger at that time.
The merger increases the density and the velocity dispersion of particles in the main halo.
The increase in velocity dispersion is particularly important, especially given that we do not see a jump in the higher $\chi_{\rm init}^2$ haloes where down-scattering is more dominant.
This is because down-scattering occurs without a velocity threshold so does not need the merger to provide a velocity boost.

\section{Conclusions}
\label{sec:conclusions}

We introduce a suite of zoom-in simulation using a SIDM model that incorporates a mix of elastic and inelastic scattering.
The model consists of a two state DM particle, and a particle can move between states during inelastic scatters by exchanging mass energy with kinetic energy.
In particular, the model includes an endothermic reaction that is just as likely as exothermic and elastic reactions at relative velocities above 400\:km/s.

Past work showed that this new model resulted in cores, with the core density set by the onset time of endothermic reactions \citep{O'Neil2023}.
Here, we investigated the effects of the initial conditions, which heavily influence this onset time.
We ran a suite of 11 SIDM simulations of the same zoom-in halo varying the initial fractions of ground and excited state particles, with the initial excited state fraction $\chi_{\rm init}^2$ between 0-100\% in increments of 10\% along with a fiducial CDM model.
We summarise our results as follows:

\begin{itemize}
    \item We show in Figure \ref{fig:density_vdisp} the density profile and velocity dispersion as a function of radius for each of our models.  The SIDM model consistently produces a core in the density profile.  The size of this core increases with $\chi_{\rm init}^2$, corresponding to more down-scattering reactions.  Similarly, the density of the inner regions decreases with $\chi_{\rm init}^2$ and the velocity dispersion also flattens out at low radii.
    The central density changes by a factor of 40 between the $\chi_{\rm init}^2=0$ and the $\chi_{\rm init}^2=1$ simulations.
    \item Figure \ref{fig:v_max-m_200} shows the maximum circular velocity as a function of $M_{\rm200,mean}$ and $\chi_{\rm init}^2$.  $V_{\rm max}$ and $M_{\rm200,mean}$ steadily decrease as a function of $\chi_{\rm init}^2$.  Lower $\chi_{\rm init}^2$ haloes are more concentrated due to increased up-scattering.
    \item For each simulation, we fit an ellipse at various radii and compare the major, middle, and minor axes to estimate the shapes of the haloes. Figure \ref{fig:axis-ratios-all} shows the middle-to-major and minor-to-major axis ratios ($q$ and $s$ respectively)at several radii for each halo, so a ratio of $s=1$ is circular and $q=0$ is flat.  Higher $\chi_{\rm init}^2$ consistently produces more oblong haloes across all radii with axis ratios as low as $q=0.6$ and $s=0.55$, whereas low $\chi_{\rm init}^2$ haloes are close to spherical.  This is due to early up-scattering compressing the halo such that it loses any oblong shape, and it then remains spherical throughout its evolution.  With early down-scattering, the halo remains puffy and there is not a strong enough influence from scattering to compress the halo into a sphere.
    Figure \ref{fig:anisotropy} shows the anisotropy parameter $\beta$ as a function of radius for each simulation, with higher $\beta$ corresponding to more radial orbits.  In the presence of self-interactions, $\beta$ decreases across all radii.
    \item The substructure is significantly decreased as $\chi_{\rm init}^2$ increases.  For the $\chi_{\rm init}^2=0$, the substructure is reduced by a factor of 2 compared to the CDM model, and the $\chi_{\rm init}^2=1$ model decreases the number of satellites by over an order of magnitude.  Figures \ref{fig:subhalo_counts_radius} and \ref{fig:subhalo_mass_function} show the number of subhaloes enclosed within a given radius and the subhalo mass function respectively.  Both of these plots show the same trend of the shape of the functions remaining consistent but moving to lower numbers.  With higher $\chi_{\rm init}^2$, down-scattering occurs frequently early in the halo's formation.  This gives particles a velocity kick, which makes it more difficult to hold them in gravitational wells.  This makes small subhaloes without much gravitational pull difficult to form, and even larger subhaloes lose some of the particles early in their evolution.  This effect occurs across radius and mass ranges, indicating that there is not a large environmental dependence on the effect but that the main effect is the initial states and how it alters the early evolution of the halo.  The exception to this is in large subhaloes $\sim10^{10}\:\msun$, where the gravity is stronger and the effect is slightly less pronounced.
    \item Figure \ref{fig:z_states} shows the fraction of excited state particles as a function of redshift for each simulation.  These lines do not cross, so if a halo starts with a higher or lower $\chi^2$ than another, this remains the case throughout the halo's evolution.  However, $\chi^2$ for each halo converge to more similar values at low redshift, so higher $\chi_{\rm init}^2$ decrease $\chi^2$ with time and lower $\chi_{\rm init}^2$ increase $\chi^2$ with time.  The haloes converge to a value that is set by the velocity dispersion of the halo as described in Equation \ref{eq:chi_eq}.
    
\end{itemize}

There are several other factors that influence halo properties that we did not consider here.
The overall environment and mergers may cause certain reactions to dominate or provide an influx of new particles and substructures.
A dedicated set of simulations using the same early universe conditions but simulating a variety of halos would shed light on the diversity problem in particular.
In addition, baryons could alter several of the properties, for example by deepening the central gravitational potential.
This could cause a difference in, for example, whether subhaloes go through core collapse or not.
Dissipative dark matter has been shown to accelerate gravothermal collapse \citep{Essig2019}, which mimics the behaviour of the endothermic reactions.
However, the two-state dark matter particle includes exothermic reactions, which causes haloes to become more diffuse.
Thus, some haloes may experience accelerated core collapse while others may not, depending on the histories of the subhaloes.
Currently, most constraints exist for elastic SIDM \citep{Adhikari2022}, while other types of SIDM are not well constrained.
Because the two-state model introduces several types of scattering, certain reactions will dominate in different situations.
Studying how the halo properties like shape and density profile change depending on mass scale and merger history could help constrain models like this one.
Larger haloes with large velocity dispersions may be more dominated by endothermic reactions since particles can more easily cross the velocity threshold.
For this, it would be interesting to run a set of cluster simulations using this model.

The initial conditions of the halo result in significantly altered halo properties.
In particular, high $\chi_{\rm init}^2$ results in a halo with properties consistent with those from exothermic SIDM, and low $\chi_{\rm init}^2$ results in properties consistent with endothermic SIDM.
To accurately predict the expected halo, it is therefore important to have a solid foundation in the initial conditions of the model.
The sensitivity of the resulting halo to the initial conditions indicate that this model can produce a range of shapes and density profiles from compact to diffuse, potentially alleviating the diversity of shapes problem.
The importance of this work lies in highlighting the sensitivity of the halo to early universe conditions while controlling for these other variables.

\section*{Acknowledgements}

MV acknowledges support through NASA ATP 19-ATP19-0019, 19-ATP19-0020, 19-ATP19-0167, and NSF grants AST-1814053, AST-1814259, AST-1909831, AST-2007355 and AST-2107724.
Simulations and analysis were done with MIT's Engaging cluster.

We made use of the following software packages in the analysis:
\begin{itemize}
	\item {\textsc{Python}}: \citet{vanRossum1995}
	\item {\textsc{Matplotlib}}: \citet{Hunter2007}
	\item {\textsc{SciPy}}: \citet{Virtanen2020}
	\item {\textsc{NumPy}}: \citet{Harris2020}
	\item {\textsc{Astropy}}: \citet{Astropy2013,Astropy2018}
	\item {\textsc{SwiftSimIO}}: \citet{Borrow2020, Borrow2021}
\end{itemize}

\section*{Data availability}
The data used in this paper are available upon reasonable request.

\bibliographystyle{mnras}
\bibliography{bibliography}

\label{lastpage}

\end{document}